# Correlated spectroscopy of electric noise with color center clusters

Tom Delord[1], Richard Monge[1], and Carlos A. Meriles[1,2,†]

**Experimental noise often contains valuable information on the interactions of a system with its environment but establishing a relation between the measured time fluctuations and relevant physical observables is rarely apparent. Here, we leverage a multi-dimensional and multi-sensor analysis of spectral diffusion to investigate the dynamics of carriers in charge traps surrounding color center clusters in diamond. Working with nitrogen-vacancy (NV) centers sharing the same diffraction-limited volume, we establish statistical correlations in the spectral changes we measure as we recursively probe the optical resonances of the cluster, which we subsequently exploit to unveil proximal traps. By simultaneously co-monitoring the spectra of multiple NVs in the set, we show the ability to deterministically induce Stark shifts in the observed optical resonances, hence allowing us to identify electrostatically coupled sets of emitters. These cross-correlated measurements allow us to determine the relative three-dimensional positions of interacting NVs in a cluster as well as the location and charge sign of proximal traps. Our results can be generalized to other color centers and open intriguing opportunities for the microscopic characterization of photo-carrier dynamics in semiconductors and for the manipulation of nanoscale spin-qubit clusters connected via electric fields.**

Often seen as detrimental, random fluctuations in the response of a probed system can nonetheless shed light on otherwise hidden physical processes. A paradigm example is Johnson noise, the intrinsic voltage fluctuations in a resistor, whose root mean square amplitude can be tied to the system temperature[1]. Another illustration can be found in the statistical fluctuations of a paramagnetic ensemble[2], where random spin coherences forming and decaying with characteristic energies and time scales allow the experimenter to reconstruct the system's magnetic resonance spectrum in the absence of resonant drives[3,4]. Further, the ability to establish correlations in the observed fluctuations brings in the resolving power of multi-dimensional spectroscopy, broadly employed in magnetic resonance[5].

While fundamental fluctuations are readily detectable in macroscopic systems[3,4], experiments at the nanoscale are often better suited for noise spectroscopy because the fractional change on the observed random signal increases as the system size reduces[6-8]. In particular, simultaneous correlations in space and time — recently proposed as a strategy to enhance the information content in quantum sensing experiments[9] — are intrinsically easier to capture in nm-sized systems.

Here we monitor the optical transitions of individual NVs within clusters sharing the same diffraction-limited volume as we randomly alter the occupation of proximal charge traps. Comparing the spectra from multiple measurements — affected by reconfiguring electric environments — we extract single- and multi-NV spectral correlations that we then use to map out the NV relative positions in three-dimensions and co-locate proximal charge traps within the crystal host. Capitalizing on the dual role of NVs — alternatively serving as an electric field probe or a carrier trap — we further illustrate how controlled ionization of an individual center from a pair in the cluster allows us to deterministically change the optical resonances of the other. Related work has been reported recently, both for NVs in diamond[10] and organic color centers in carbon nanotubes[11].

## NV electronic structure and experimental design

Throughout our experiments, we use narrow-band, tunable laser excitation (637 nm) to monitor small sets of negatively charged nitrogen-vacancy (NV) centers in diamond via confocal microscopy. Formed by a substitutional nitrogen immediately adjacent to a vacancy[12], these spin-active color centers are presently attracting broad attention as a platform for quantum information processing[13] and nanoscale sensing[14]. At low temperatures, the zero-phonon line features a fine structure, best captured through the energy diagram in Fig. 1a: Importantly, the $^3E$ excited state manifold is highly sensitive to the crystal environment[12], resulting in a set of optical resonances whose exact frequencies depend on the transverse and longitudinal components of the local electric and strain fields (respectively, $\delta_\perp$ and $\delta_\parallel$). Negative to most applications[13], this NV susceptibility is instead valuable herein, as it heralds physical reconfigurations in the occupation of proximal charge traps (schematic in Fig. 1a).

---

[1]Department. of Physics, CUNY-City College of New York, New York, NY 10031, USA. [2]CUNY-Graduate Center, New York, NY 10016, USA. [†]E-mail: cmeriles@ccny.cuny.edu

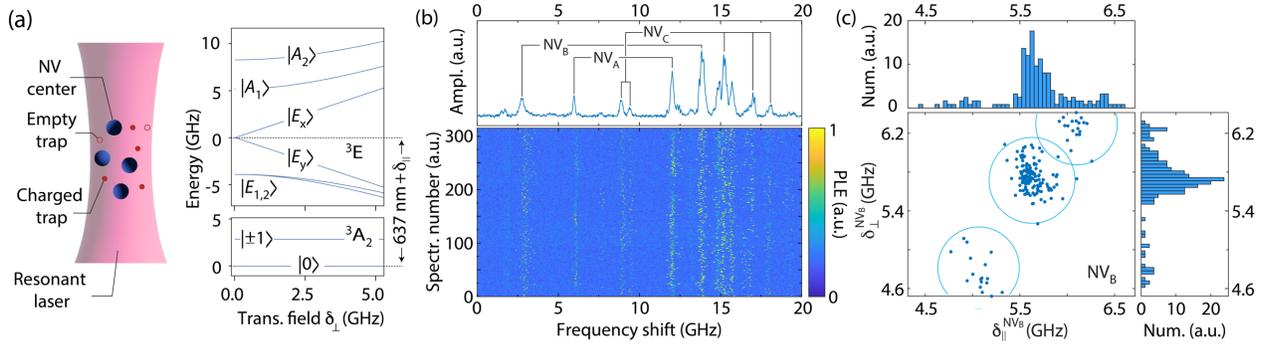

**Figure 1 | Optical spectroscopy of diffraction-limited NV clusters.** (a) (Left) We use resonant confocal microscopy to individually address NVs from a small set sharing the same diffraction-limited volume. (Right) Under cryogenic conditions, the NV⁻ excited state manifold ³E splits into two triplets yielding a collection of optical resonances around 637 nm whose values depend on the electric and strain fields at each NV site. (b) Recursive photo-luminescence excitation (PLE) spectroscopy of a cluster featuring three different NVs. The bottom plot shows successive spectra under simultaneous MW (2.88 GHz) and tunable 637 nm light preceded by charge initialization with 532 nm light. The upper 1D plot is the integrated sum of all individual traces; spectral-diffusion-induced broadening of all resonances is apparent. (c) Correlated longitudinal and transverse fields on NV_B as extracted from the spectra in (b). Data clustering (indicated by circles) is apparent in the 2D plot but not so much in the 1D projected histograms (upper and right inserts). In (b), the green (red) laser power during charge initialization (frequency sweep) is 1 μW (3 nW) and the reference frequency is 470.000 THz. Unless otherwise noted, all experiments are carried out at 7 K in the absence of any externally applied magnetic field.

We introduce our working strategy in Fig. 1b where we plot a series of photo-luminescence excitation (PLE) spectra from a first NV cluster. Combined use of selective NV⁻ ionization and microwave excitation allows us to identify three NVs within the same diffraction-limited volume (hereafter denoted A, B, and C, see also Supplementary Material, Sections 1 through 4). Spectral 'diffusion' of the optical resonances is apparent as we intercalate pulses of green light (532 nm) between successive sweeps of the tunable laser. The power and duration of these pulses — adjusted to cycle the NV charge state between negative and neutral — also lead to concomitant changes in the occupation of proximal traps, hence resulting in a varying electric environment.

While fluctuations in the occupation of the many traps far from an NV induce a near-continuous diffusion of the optical spectrum, proximal charge state changes must arguably lead to discernible spectral jumps. Further, the energy diagram in Fig. 1a ties the optical spectra to $\delta_\perp$ and $\delta_\parallel$ (respectively associated to shifts between optical resonances, and of the entire multiplet[15]), which can be leveraged to extract information not apparent when considering a time-averaged spectrum (upper insert in Fig. 1b). We first demonstrate this notion in Fig. 1c where we build on the spectra measured for NV_B to derive a two-dimensional (2D) field histogram correlating the values derived for $\delta_\perp$ and $\delta_\parallel$. We find a highly non-uniform distribution, which hints at discrete jumps along a quasi-Gaussian electric noise background (see below). Crucially, data bunching is far less obvious if one only considers each field projection separately (upper and right inserts in Fig. 1b) or, more directly, the one-dimensional (1D) probability distributions of individual optical resonances, which underscores the need for a correlated, multivariate analysis.

**Mapping out nearby charge traps**

To better illustrate the advantages and limitations intrinsic to this class of noise spectroscopy we resort to NV_C where the 2D field histogram of $\delta_\perp$ and $\delta_\parallel$ reveals a more complex, though richer structure (Fig. 2a). To interpret our observations, we model the electric field environment as the stochastic sum of contributions from carriers at three proximal sites (Figs. 2b and 2c). Assuming each trap intermittently hosts one fundamental charge, we converge to a set of four "primary" configurations corresponding to states where all traps are empty, or where a carrier occupies one of the three possible sites.

One can now combine these alternative scenarios to predict the transverse and longitudinal fields expected in cases where more than one trap is occupied. We demonstrate this notion in Fig. 2b where we add the fields produced by carriers in traps 1 and 2 (respectively, $(q_1, 0, 0)$ and $(0, q_2, 0)$ in Fig. 2b) to obtain the data subset in the upper right corner of the plot, $(q_1, q_2, 0)$; it is easy to see the same applies to the two alternative cases (sites 1 and 3 or 2 and 3, faint vectors in Fig. 2b). Since transverse components — only determined in magnitude — follow correctly from a vector sum, we conclude that all traps must approximately lie on the same plane; further, because each individual trap hosts on average one carrier $(13\pm2)\%$ of the time (Fig. 2a), we expect to find two simultaneously full traps with a probability of only $(2\pm1)\%$, in good

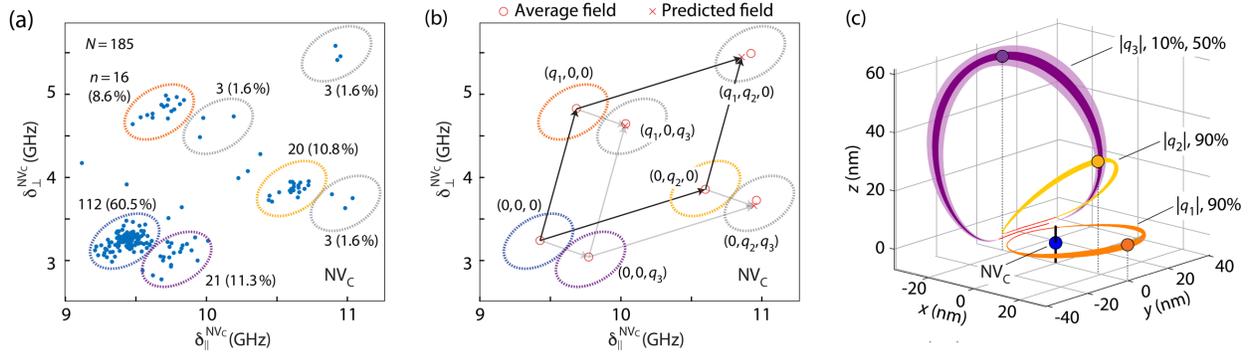

**Figure 2 | Mapping out proximal charge traps.** (a) Longitudinal and transverse fields for $NV_C$ as extracted from the spectroscopy set in Fig. 1b; $N$ indicates the total number of observations, and $n$ is the number of data points enclosed in each ellipse. (b) We model the electric field on $NV_C$ as the combined effect from point charges in three proximal traps and a fluctuating field environment of more distant carriers. The probability of a given charge configuration can be extracted from the fractional weight of each data cluster in the plot where not more than one trap is populated. This information is sufficient to predict the fields and occurrence probability of those configurations where more than one trap is occupied (see crosses within gray ellipses). (c) Probability isosurfaces for the point charge positions yielding the average fields shown in (b) for the case of electron-populated traps; percent values denote confidence intervals. Red lines indicate sections of otherwise possible solutions, here discarded based on observations with two simultaneously populated traps.

agreement with our observations ($(1.6\pm1)$%, grey ellipses in Figs. 2a and 2b).

Unfortunately, changes in the transverse and longitudinal fields as measured from a single NV are insufficient to completely determine the position and charge of a trap. In addition, our measurement uncertainty creates a spread in space for the probability distribution of the trap's position (see Supplementary Information, Sections 5 and 6). We plot in Fig 2c the isosurfaces of such probability distributions calculated for the trap set probed by $NV_C$ (Figs. 2a and 2b), assuming negative charges. The presence of a background bias field — combining the local strain and residual electric fields from traps not ionized by green illumination, see Supplemental Information, Section 7 — makes the shapes of these curves non-trivial. Interestingly, we can exclude sections of the solution set incompatible with observations where more than one trap is populated (red segments in each loop in Fig. 2c). We show below this form of "co-measurement" can be adapted to cases where more than one NV picks up the field from a single trap, an approach that dramatically increases the information on both the location and sign of the captured carrier.

**Correlated spectroscopy across multiple color centers**

Optical excitation is known to cycle the NV charge state between neutral and negatively charged[16] — respectively, $NV^-$ and $NV^0$ — implying that NVs alternatively act as local probes or as point sources of electric field. A key implication is that a change in the charge state of a given NV must correspondingly lead to observable spectral shifts in their neighbors. We illustrate this idea in Figs. 3a and 3b where we study a new sub-diffraction-size cluster comprising four NVs (here labeled with letters D through G). The green laser pulse we apply prior to the optical frequency sweep randomly initializes the charge state of each NV into negative or neutral, hence making them observable or not upon resonant excitation (the zero-phonon lines of $NV^0$ lie far away from the scanning range of the tunable laser[17]). Specifically, the yellow arrows on the right-hand side of Fig. 3b highlight instances where the $NV_G^-$ resonances are missing, indicative of green-induced ionization (typically occurring with ~20% probability[16,18]). We observe in each case a change in the relative peak amplitudes of $NV_F^-$ as well as a strong blueshift of its resonance frequencies (orange arrows); indeed, we conclude from the field histogram in Fig. 3c that the correspondence of these shifts with a direct optical read-out of charge state of $NV_G$ is nearly perfect.

The use of narrow-band excitation gives us the opportunity to gain control on the spectral response upon selective NV ionization. We validate this idea in Fig. 3d, where we first use weak red light to post-select an instance in which both $NV_F$ and $NV_G$ are negatively charged (blue trace). We subsequently tune the laser frequency and amplitude to ionize only $NV_G$[18]; following a second spectral sweep, the concomitant spectral changes we induce in $NV_F$ become apparent (red trace). Note that the converse experiment — where we probe $NV_G$ upon selectively ionizing $NV_F$, Fig. 3e — yields analogous phenomenology, although the magnitude of the frequency shift is substantially different. This asymmetry reflects the different orientations of the transverse background bias fields acting on each NV.

The longer range of Coulombic couplings hints at a fully interacting cluster, where altering the charge state of one of the traps reverberates on all others. Exposing the effect of more weakly coupled NVs, however, becomes challenging as the spectral diffusion caused by many coexisting traps becomes dominant. We circumvent this complication in Fig. 4a where we cross-correlate the $\delta$-fields acting on $NV_D$ and $NV_F$. We find that the charge state of $NV_D$ — here serving as a third classifier, blue and red dots — has an impact not only on $\delta_\parallel^{NV_F}$ but, more importantly, on the mean values of $\delta_\parallel^{NV_D}$ and $\delta_\perp^{NV_D}$, a response that reveals $NV_D$'s coupling to $NV_G$ (we find a similar response in the case where we exchange the roles of $NV_F$ and $NV_G$, hence allowing us to conclude $NV_D$ also couples to $NV_F$).

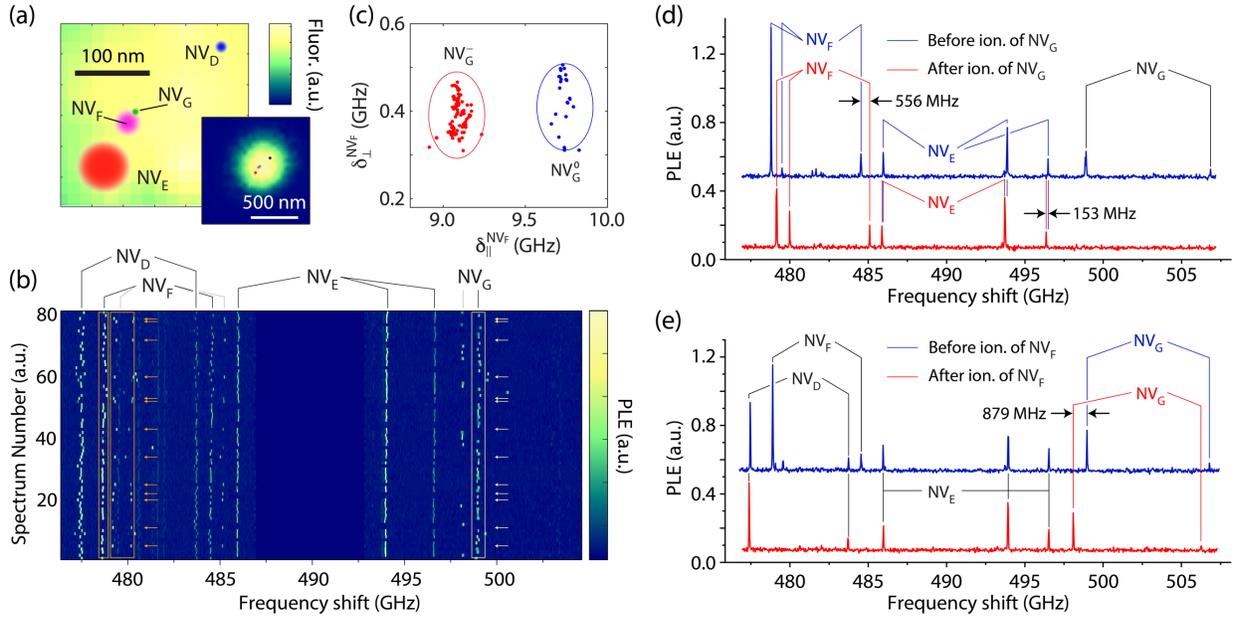

**Figure 3 | Coulomb-field control of NV optical resonances.** (a) Confocal image (532 nm excitation) of a second cluster comprising four different NVs. Superimposed circles indicate the in-plane positions as determined from confocal imaging using laser light resonant with the $E_y$ transition of each NV in the set; the radius in each circle indicates the uncertainty. The lower right insert is a zoomed-out image of the same cluster. (b) Recursive optical spectroscopy of the NV cluster in (a). Individual inspection of the spectra in the series reveals a correlation between $NV_G$ ionization (heralded by a missing optical resonance, yellow arrows) and the shift to longer wavelengths in the optical resonances of $NV_F$ (orange arrows). All conditions as in Fig. 1b. (c) Correlation spectroscopy between the longitudinal and transverse fields at $NV_G$ as derived from the results in (b). Red and blue dots respectively denote instances where $NV_D$ is negative or neutral; ellipses are guides to the eye. (d) Starting from a configuration where both $NV_F$ and $NV_G$ are negatively charged (blue trace), we probe the cluster response after resonant ionization of $NV_G$ (red trace). (e) Same as in (d) but for $NV_F$ ionization. The reference frequency in (b) and (d) is 470.000 THz.

Unlike green illumination, weak red excitation has a reduced effect on the charge states of NVs and most traps, hence suggesting the use of time correlations between consecutive spectral sweeps as a second alternative to mitigating spectral diffusion. Figure 4b shows the field histogram of $NV_D$ as well as the difference between the fields measured via successive spectral sweeps in the absence of NV charge initialization by green light. This time-correlation measurement effectively suppresses slow background spectral diffusion, and dramatically sharpens clusters created by light-induced changes in the charge states of $NV_F$ and $NV_G$ (Supplementary Material, Sections 7 and 8). The flipside is a longer experimental time and a partial loss of information as the population statistics of a trap cannot be determined. Worth noting, a similar analysis on $NV_E$ shows much weaker trends suggesting this color center is far from the rest (our diffraction limited volume stretches 1–2 μm along beam axis).

We can now combine the information gathered thus far to map the NV cluster in three dimensions (Fig. 4c). In short, we use a search algorithm that tests all positions of the NVs with respect to each other, using measurements from the field histograms of $NV_F$ and $NV_G$ to calculate the relevant probability distributions (Supplementary Information, Section 9). Figure 4c illustrates such a test performed for the most likely positions (spheres), showing perfect intersection between independent measurements of $NV_D$ (cyan and purple loops) and $NV_F$ (green and magenta). We estimate the relative distances of $NV_F$ and $NV_D$ relative to $NV_G$ respectively as $(48 \pm 1.6)$ nm and $(150 \pm 23)$ nm, with the greater uncertainty arising from the weaker couplings.

Recurrent optical spectroscopy over an extended data set shows simultaneous shifts of $NV_F$ and $NV_G$ approximately 2% of the time, which we interpret as a manifestation of intermittent carrier capture by a trap proximal to both NVs. Here too, we leverage the set of solutions extracted from either NV to co-locate the trap position, approximately $(29 \pm 2)$ nm from $NV_F$ $((60 \pm 3)$ nm from $NV_G$, Fig. 4d). Note the probability distributions do not fully intersect, likely a consequence of underestimated error sources during data analysis. Interestingly, changing the nature of the trapped carrier from an electron to a hole leads to clearly disjoint solution sets and can be ruled out (fainter loops in Fig 4d). Co-localization of a trap, therefore, lifts ambiguities in the captured charge sign (and amplitude) inherent to single NV probe sensing, in the process providing clues on the physical nature of the trap: Potential candidates include a substitutional nitrogen impurity (here transitioning from $N^+$ to $N^0$) or a lattice vacancy (sporadically changing from $V^0$ to $V^-$).

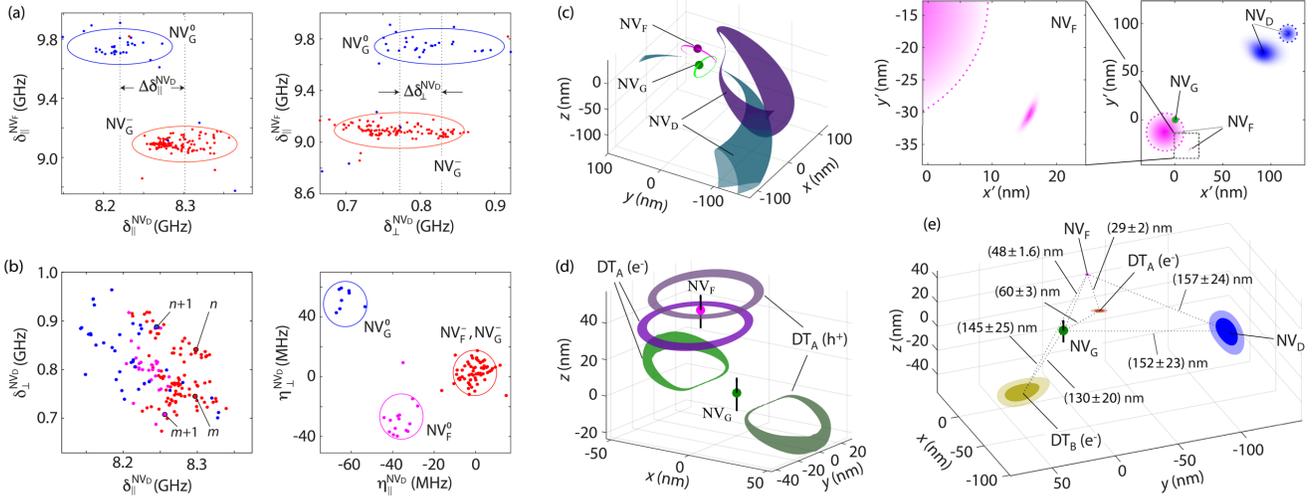

**Figure 4 | 3D co-localization via multi-NV cross-correlated noise spectroscopy.** (a) Correlations across $NV_D$ and $NV_F$ as derived from Fig. 3b; a charge state change of $NV_G$ (red or blue dots) leads to a shift in the mean electric field projections at $NV_D$; ellipses are guides to the eye. (b) (Right) Histogram of the electric field as seen by $NV_D$; blue (magenta) dots indicate instances where $NV_G$ ($NV_F$) is neutral, while red indicates no ionization. Integers $n, m$ highlight examples corresponding to time consecutive spectra. (Left) Same as before but calculated as δ-field differences $\eta = \delta_{n+1} - \delta_n$ between two successive measurements; the suppression of slower temporal fluctuations leads to higher spectral resolution. (c) Co-localization of NVs in the cluster of Fig. 3. The magenta (green) loop represents the probability distributions for $NV_F$ ($NV_G$) as seen by $NV_G$ ($NV_F$). Similarly, the purple (cyan) areas represent probability distributions for $NV_D$ using $NV_G$ ($NV_F$) as the probe; confidence intervals are 70% for $NV_D$ and 95% for the rest. (d) Colocalization of dark trap $DT_A$, proximal to $NV_F$ and $NV_G$; upper and lower loops show the 90% probability distributions for the position of the trap as seen by $NV_G$ and $NV_F$, respectively. (e) (Main) Three-dimensional spatial locations of $NV_D$, $NV_F$, $NV_G$, and proximal traps $DT_A$, $DT_B$; confidence intervals from dark to light are 50% and 90%. (Upper inserts) Spatial locations of $NV_D$, $NV_F$, $NV_G$ projected on the $xy$ optical plane as extracted from the 3D plot. The dashed circles show the NVs location measured from super-resolution (95% CI); all locations are relative to $NV_G$.

Figure 4e integrates the findings above into a combined 3D plot that includes all three interacting NVs in the cluster and dark trap $DT_A$, as well as a second trap, $DT_B$, observed to capture an electron with a probability — 19% on average — that seems to depend on the charge state of $NV_F$ (Supplementary Material, Section 10); this occupancy is much higher than that of $DT_A$ (2%), and hints at a different physical nature. To compare the cluster geometry against that derived from optical imaging, we leverage the known NV orientations to determine the transformation matrix connecting the $xyz$ frame in the figure to the $x'y'z'$ reference frame in the lab. As shown in the upper inserts, projecting the calculated positions for

$NV_D$, $NV_F$, and $NV_G$ onto the $x'y'$-plane — perpendicular to the incoming laser beam — yields an image consistent with that attained via super-resolution microscopy.

## Conclusions and outlook

Statistical correlations in the spectral noise affecting the optical resonances of sub-diffraction NV clusters contain valuable information on the underlying electrostatic fluctuations of the crystal environment. Specifically, we exploit otherwise detrimental spectral diffusion in the optical response of an NV to single out proximal charge traps and set bounds on their physical locations. Further, we build on the ability to resonantly ionize $NV^-$ to expose Coulombic couplings between the NVs of a cluster — even in the presence of substantial environmental noise — which then allow us to determine their relative positions in three dimensions. Central to these findings is the notion of co-measurement by two (or more) atomic size probes, which we apply herein to map out the position of a remote color center in a cluster of several, and to pinpoint the location of a non-fluorescent charge trap as well as the magnitude and sign of the carrier it captures.

While NV centers are key to the present findings, most ideas can be extended to other color centers in diamond or alternative material hosts, provided a known relation exists between the observed optical resonance spectrum and the electric environment[19,20]. Relevant examples include the silicon vacancy[21] ($V_{Si}$) and carbon-silicon di-vacancy[22] ($V_{Si}V_C$) in SiC, as well as group-IV vacancy color centers in diamond[23-26] (which can be sensitive to proximal traps despite their symmetry-based first-order protection against electric fluctuations[27]). From a methodological point of view, one can envision extensions in the form of protocols adapted to investigating the trap response under optical excitation not affecting the NV charge state or, alternatively, to probing the diffusion of electrons injected from proximal NVs serving as a source[28,29].

Our approach paves the route to a deeper understanding of the microscopic mechanisms underlying spectral diffusion, and hence promises opportunities to developing novel schemes for electric field sensing[30], or for tasks in quantum information processing relying on indistinguishable photons[31]. Along related lines, the comparatively long range of Coulombic couplings could be exploited to mediate interactions between spin qubits otherwise too far from each other to couple magnetically[32,33]. Besides applications in quantum science and technology, future material science studies will be required to shed light on the formation of NV clusters in bulk diamond — observed in our crystals with unanticipated frequency — as an intriguing alternative to implanting nitrogen-rich moieties[34].

## Data availability

The data that support the findings of this study are available from the corresponding author upon reasonable request.

## Code availability

All the MATLAB source codes for data analysis and 3D reconstruction used in this study are available from the corresponding author upon reasonable request.

## Acknowledgments

We acknowledge Olaf Bach for fruitful discussions and assistance with some of the modeling. C.A.M acknowledges support from the National Science Foundation through NSF-2216838; T.D. acknowledges support by the U.S. Department of Energy, Office of Science, National Quantum Information Science Research Centers, Co-design Center for Quantum Advantage (C2QA) under contract number DE-SC0012704. R.M. acknowledges support from grants NSF-1914945 and NSF-2316693. All authors acknowledge access to the facilities and research infrastructure of the NSF CREST IDEALS, grant number NSF-2112550.

## Author contributions

T.D., R.M., and C.A.M. conceived the experiments. R.M. and T.D. conducted the experiments. T.D. and R.M. analyzed the data with C.A.M.'s assistance. C.A.M. supervised the project and wrote the manuscript with input from all authors.

## Competing interests

The authors declare no competing interests.

## Correspondence

Correspondence and requests for materials should be addressed to C.A.M.

## Supplementary Information

### 1. Experimental

We make use of a custom-made confocal microscope integrating a narrow-linewidth (500 kHz), 637-nm laser tunable across a 4 nm range (Toptica DL PRO), and two off-resonant, continuous-wave (cw) 532-nm lasers. We combine all light sources into a single-mode fiber and use a 605-nm dichroic mirror and a 70/30 beam splitter to selectively collect the sample fluorescence; we control the sample temperature via a closed-cycle cryo-workstation from Montana Instruments[18,35]. Acousto-optic modulators (AOM) in a double-pass configuration produce 10-ns-risetime laser pulses for the 637-nm laser as well as one of the 532-nm lasers (Coherent Sapphire 532-100 CW SF).

To avoid green bleed-through during a low-power red scan, we replace the AOM-controlled 532-nm laser by a 532-nm laser diode (Thorlabs DJ532-40), which we then use for NV charge initialization (with typical power of 200 μW). We carry out NV excitation and photoluminescence (PL) collection via a 0.75-NA Zeiss objective sitting on a room temperature (RT) platform within the cryo-workstation main chamber; all experiments are performed at 9 K. Two signal generators (Rhode & Schwarz and Stanford Research Systems) serve as the source of microwave (MW), which we deliver through a 25-μm wire overlaid on the sample; we use switches from Minicircuits to produce MW pulses, and implement time-resolved protocols with the help of a pulse generator (SpinCore's Pulse-Blaster).

Throughout our experiments, we study two electronic-grade, [100] diamond crystals purchased from Element 6[36,37] and Delaware Diamond Knives[37], both with approximately 5 ppb of nitrogen content. We probe naturally occurring NV clusters ~5 μm deep in the bulk.

### 2. Determining the NV spatial orientation

To determine the orientation of all NV's with respect to the laboratory frame, we make use of a permanent magnet outside the cryo-workstation to create a weak magnetic field $B$ at the sample site. The field orientation was determined precisely from the optically-detected magnetic resonance (ODMR) spectra of three crystallographically inequivalent NVs[38]. For the NVs of a cluster, we carry out selective ODMR measurements via resonant red readout[39]. Combining the observed frequencies with the measured field magnitude and orientation, we extract the physical lab frame orientation of each NV resorting to known protocols[38,40]. Relevant to the results in Fig. 4 of the main text, we find that $NV_E$, $NV_F$, and $NV_G$ share the same orientation, while $NV_D$ points along a different crystallographic axis.

### 3. PLE measurements and selective NV⁻ ionization

We implement two alternative methods: The first one — referred to as "repump PLE"[41] — alternates a green pulse (532 nm, 1 mW, 1 μs, Coherent Sapphire) and a red pulse (637 nm, 500 nW – 4 μW, 500 ns), respectively used to initialize the NV charge state and subsequently probe the NV fluorescence at a given wavelength (Supplementary Fig. 1a). For each laser wavelength in the scan, we typically repeat this protocol $10^5$ times; note that green illumination not only impacts the NV charge but also the charge state of the environment, implying the resulting PLE spectrum amounts to an average over all possible configurations.

Supplementary Fig. 1b lays out the second approach, here referred to as "repump-free PLE". This protocol starts with a single green pulse (532 nm, 200 μW, 1–10 s, Thorlabs DJ532-40), which we use to charge-initialize the NV cluster and spatially track the photoluminescence (thus accounting for long term drifts). We then apply a weak readout pulse (637 nm, 3–10 nW) simultaneously with continuous MW excitation at 2.877 GHz for 1.2–2.4 seconds per wavelength step, absent of any applied magnetic field. The entire protocol takes 10-20 minutes for a ~400 point spectrum.

The experiments in Fig. 3 of the main text rely on selective ionization of a target NV⁻ in the cluster, which we implement via an adapted version of the repump-free PLE protocol (Supplementary Fig. 2a). Specifically, we first determine the NV charge state via a non-destructive PLE scan using only 3 nW of 637-nm excitation (Supplementary Fig. 2b). We subsequently stabilize the red laser at a wavelength resonant with one of the optical transitions of the target NV, to then monitor the resulting fluorescence under continuous illumination; a sudden signal drop flags ionization (Supplementary Fig. 2c). Lastly, we confirm the procedure acted selectively on the target NV via a second non-destructive PLE scan (Supplementary Fig. 2d).

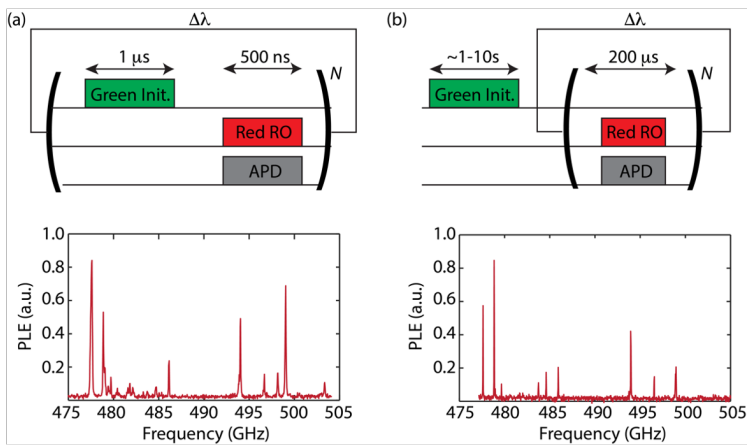

**Supplementary Figure 1: PLE acquisition protocols.** (a) (Top) Schematics of "repump" PLE. (Bottom) Example NV cluster spectrum upon application of the above protocol. (b) Same as in (a) but for a "repump-free" protocol. By comparison, the absence of charge initialization via a green reset at each wavelength leads to considerably narrower lines, which, nonetheless become subject to spectral diffusion upon repeated observations. In (a) and (b), $\Delta\lambda$ represents the wavelength step during a laser scan, and $N$ is the number of repeats at a given wavelength.

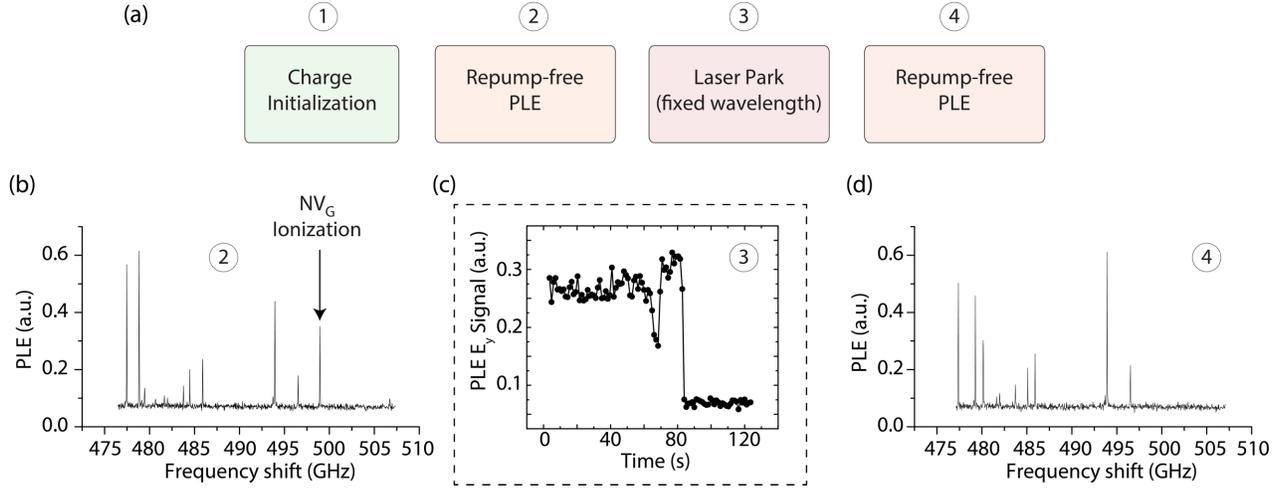

**Supplementary Figure 2: NV-selective ionization.** (a) Schematics of the protocol. (b) PLE spectrum from the NV cluster in Fig. 3 of the main text prior to selective ionization. The arrow points to the chosen $NV^-$ optical resonance, in this case the $E_y$ transition of $NV_G$. (c) Amplitude of $E_y$ resonance as a function of time; $NV^-$ ionization leads to a sharp drop. (d) Repump-free spectrum of the same NV cluster upon selective ionization of $NV_G$.

## 4. PLE assignments and sub-diffraction NV imaging

To establish a relation between the NV eigen-energies and the local strain/electric field $\delta$, we use the expression for the NV excited state Hamiltonian, $H_{es}$, derived by Rogers et al.[42] Upon numerically diagonalizing this Hamiltonian, we find the $\delta$-dependent eigenvalues presented in the upper plot of Fig. 1a in the main text. To assign each peak in a PLE spectrum, we adapt a repump PLE protocol so as to include MW inversion pulses resonant with the crystal field splitting of the ground state triplet; reversal of the $NV^-$ spin population from $m_S = 0$ — the spin projection after green excitation — to $m_S = \pm 1$ allows us to identify the $E_x$ and $E_y$ resonances via a drop in their PLE amplitudes; with $\delta_\perp = \left(\delta_x^2 + \delta_y^2\right)^{1/2}$ determined from the frequency splitting between these two resonances, all other assignments follow immediately from the set of eigenvalues derived for $H_{es}$. Note that the expression above assumes $\delta_\parallel = 0$; a non-null value in a measurement amounts to a net shift of all spectral lines relative to a reference.

When working with clusters, we resort to different strategies to confirm the assignments we have made. For example, we implement selective ionization protocols to identify the subset of resonances in the cluster spectrum corresponding to an individual NV within the set. Further, we check the line assignments of a given NV by monitoring the cluster spectrum upon MW-assisted repump PLE: Note that while the $E_x$ and $E_y$ PLE amplitudes fall off upon depleting the $m_S = 0$ spin projection, all others — i.e., $E_1$, $E_2$, $A_1$, and $A_2$ — must grow as they are preferentially associated to $m_S = \pm 1$ spin states.

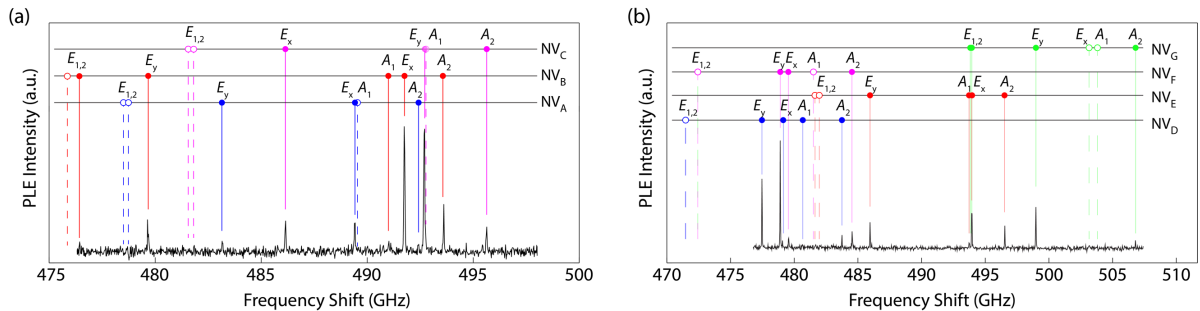

**Supplementary Figure 3: Assignment of optical resonances.** (a) Repump-free PLE spectrum for the NV cluster in Fig. 1 of the main text under conditions similar to those in that figure. Assignments for each NV follow from combining selective ionization and MW pulse excitation as described in the text. (b) Same as in (a) but for the NV cluster in Fig. 3 of the main text. In (a) and (b), the reference frequency in 470.000 THz.

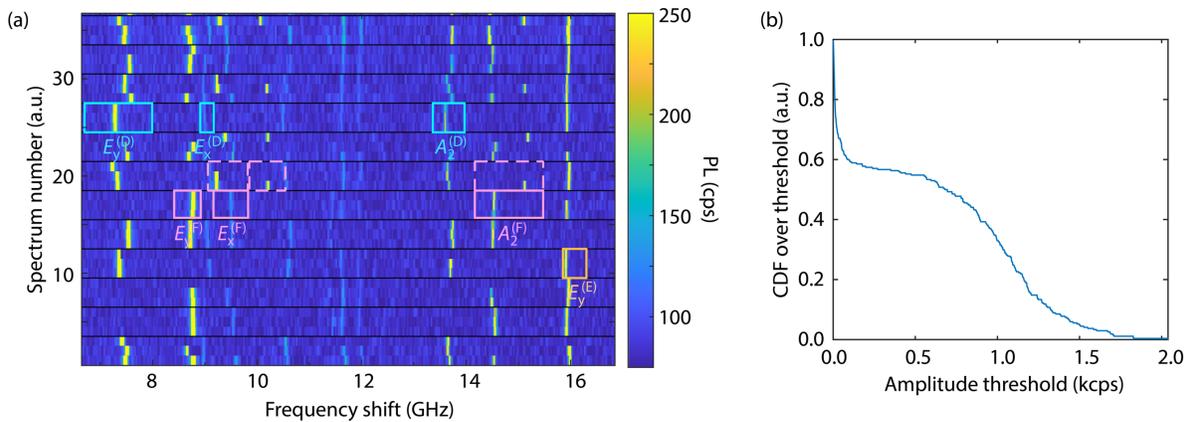

**Supplementary Figure 4: Automated analysis of a spectral series.** (a) Spectral series for the cluster in Fig 3 of the main text (low frequency end only) illustrating some of the "bands" we use in our code during automatic searches of a given resonance. Colored boxes of indicate different NVs in the cluster. Dotted rectangles indicate secondary spectral bands in $NV_F$ chosen so as to take into account the shift introduced by sporadic ionization of $NV_G$. Black horizontal lines every 3 spectra denote resets of the NV charge state with green light. The resonant laser power is 3 nW and the PL acquisition time per frequency step is 2 s; the number of points in each spectrum is 700 amounting to an acquisition time of 25 minutes per scan. Frequency shifts are referenced to 470.470 THz. (b) Cumulative distribution function (CDF) upon a Gaussian fit of $E_x^{(F)}$ in the first spectral band (solid pink rectangle). For an amplitude threshold of 0.075 kcps (set by default in our analysis code), we find a PLE line in 61% of the spectra.

We determine the spatial positions of all NVs in a sub-diffraction cluster by implementing confocal imaging under resonant excitation[18]. To this end, we first tune the laser frequency to one of the optical transitions pertaining to the NV of interest (assumed known after proper assignment), and then implement the repump PLE protocol as we galvo-scan the laser across the field of view. We finally determine the NV position with sub-diffraction resolution from a two-dimensional point-spread Gaussian fit of the resulting image.

## 5. Automated analysis of multiple consecutive spectra

We process all optical spectra in a series via a custom-designed algorithm, the purpose of which is to determine throughout a given frequency scan the spectral positions of at least two optical resonances (including $E_x$ and $E_y$) for each NV in a cluster. A prerequisite in implementing this protocol is the manual assignment of all resonances (see Section on PLE assignment above), which we then leverage to properly configure our code for automated identification across a spectral series. In particular, we rely on the early assignments and thorough visual inspection of the spectral series to preset "search bands" for each resonance, i.e., spectral windows where we anticipate finding a PLE line (Fig. S3a). Note that for NVs known to experience large discrete jumps (e.g., $NV_F$ and $NV_G$ in the cluster of Fig. 3 in the main text), more than one band might be assigned to a given transition (see solid and dashed pink rectangles in Supplementary Fig. 4a).

Whenever two resonances (often belonging to two different NVs) overlap, we make use of other, more isolated optical transitions of the same NV to narrow down the likely frequencies (e.g., we use the strong correlation between the $A_2$ and $E_x$ resonances to identify the approximate frequency of transition $E_x$, even when proximal to a resonance from another NV). We extract the central frequency, width, and amplitude of each PLE line from a Gaussian fit, which we activate only when the observed PLE amplitude exceeds a predefined threshold (Supplementary Fig. 4b).

For a given set of NV resonances, we determine $\delta_\perp$ from the set of eigenenergies derived from diagonalizing $H_{es}$. In principle, any one pair of two lines suffices, but some pairs are more sensitive (e.g., $E_x$, $E_y$) than others (e.g., $A_1$, $A_2$). When relying on more than two resonances, we ensure optimal accuracy by first calculating the electric field from every combination of PLE lines, and subsequently determining an average weighted by the derivative of the extracted electric field relative to the frequency difference between the corresponding resonances. For each NV in a cluster, we use the same number of resonances — 2 to 4 depending on the NV and experiment — to determine the electric fields for all spectra in the series.

While $\delta_\perp$ relates to the *frequency differences* between resonances, we extract $\delta_\parallel$ by comparing the *average frequency* in each NV spectrum to a fixed reference; note this reference can be arbitrary as our measurements are only susceptible to electric field changes (i.e., we cannot determine the absolute number of elementary charges in a trap). Further, charges of opposite signs lead to different spatial probability distributions, hence creating an ambiguity that can only be lifted by a co-analysis of the field fluctuations experienced by a second probe NV. We return to this point below.

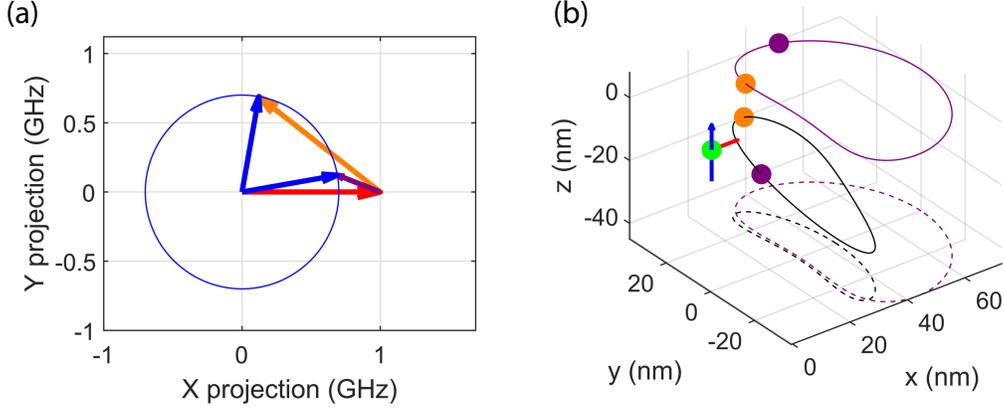

**Supplementary Figure 5: Illustrating the impact of an underlying bias field.** (a) Effect of the bias field (red arrow) on the measurement of the transverse field produced by an added charge. Since a measurement determines only the magnitude $\delta_\perp$ (blue circle), different fields (e.g., purple and orange arrows) can both lead to the same end observation despite their different magnitude and orientation (blue and light-blue arrows). (b) Effect on the localization of a discrete charge trap. The purple and orange circles are the positions corresponding to the purple and orange arrows in (a), matched by color. The purple (black) line shows all possible positions for a negative charge to generate an end transverse field with the magnitude shown in (a) without (with) an identical change of longitudinal field (300 MHz). In this schematic, the green dot represents the NV at the origin, the red arrow points in the direction of the bias transverse field, and the blue arrow represents the direction of the NV crystal axis.

When analyzing the impact of a given NV (later referred to as the "source" NV) on another one in the cluster (the "probe" NV), we rely on the presence or absence of a reference optical line in the source NV spectrum to flag its charge state during a PLE sweep. Since spectral scans proceed from lower to higher frequencies, we choose this reference as the highest-energy optical transition in the source NV spectrum so as to mitigate errors arising from mid-scan ionization.

Throughout the manuscript, we express $\delta$-fields (i.e., strain/electric fields) in GHz. To convert between units, we use the proportionality factor $\mu_E = 6.3 \frac{\text{GHz}}{\text{MV m}^{-1}}$, corresponding to a dipole moment difference $\Delta\mu = 1.3$ Debye between the NV$^-$ ground and excited states[43]. Interestingly, this value is not known with accuracy[12,44-46], and could be re-derived by combining the present technique with a super-resolution measurement of two NVs in the same plane.

## 6. Charge trap mapping via a single NV

Let us consider an individual charge trap near a probe NV and assume we can measure its charge state. By studying the NV field histogram, we can calculate the average magnitude of the electric field transverse and longitudinal components at the NV site depending on the charge state of the trap, which allows us to set boundaries on the trap's spatial location. Since we only determine the transverse field magnitude and not its orientation, the angle $\theta$ describing the transverse electric field rotation upon an electric field change remains undetermined, implying the solution takes generically the form of a closed loop in 3D space (Supplementary Figure 5).

More formally, let $\mathcal{E}_\parallel$, $\mathcal{E}_\perp$ (respectively, $\Upsilon_\parallel$, $\Upsilon_\perp$) be the longitudinal and transverse fields with (without) an added charge $q$. Assuming the NV sits at the origin, we find that for a given rotation $\theta$ of the transverse field, the charge trap must be located at a position $R(\theta, \mathcal{E}_\parallel - \Upsilon_\parallel, \mathcal{E}_\perp, \Upsilon_\perp)$ given by

$$\boldsymbol{R} = \left(\frac{q}{4\pi\epsilon_d}\right)^{1/2} \frac{\mathcal{E}_c}{|\mathcal{E}_c|^{3/2}}, \tag{1}$$

Where $\epsilon_d$ is the dielectric constant of diamond and $\mathcal{E}_c$ denotes the field due to $q$, here expressed as

$$\boldsymbol{\mathcal{E}_c} = \begin{pmatrix} \mathcal{E}_\perp \cos(\theta) - \Upsilon_\perp \\ \mathcal{E}_\perp \sin(\theta) \\ \mathcal{E}_\parallel - \Upsilon_\parallel \end{pmatrix}. \tag{2}$$

The above formula for the trap position $\boldsymbol{R}$ is only valid for exact measurements, and we must in practice deal with the uncertainty created by other fluctuations of the field or by experimental noise. We now focus on calculating the conditional probability distribution $P(\boldsymbol{r}|\boldsymbol{M})$ of finding a charge at position $\boldsymbol{r}$ given a measurement $\boldsymbol{M} = (\Delta\mathcal{E}_\parallel, \mathcal{E}_\perp, \Upsilon_\perp)$ with variance $\boldsymbol{\sigma_M^2} = (\sigma_{\Delta\mathcal{E}_\parallel}^2, \sigma_{\mathcal{E}_\perp}^2, \sigma_{\Upsilon_\perp}^2)$, where $\Delta\mathcal{E}_\parallel = \mathcal{E}_\parallel - \Upsilon_\parallel$, and $\sigma_\nu^2$ denotes the variance for variable $\nu$. We follow two alternative routes to calculate $P(\boldsymbol{r}|\boldsymbol{M})$, each featuring complementary computational speed and accuracy. Our first strategy determines the most likely position of the trap following a Bayesian approach. Let us assume a single charge is within a certain volume $V$ of the NV center. A measurement $\boldsymbol{M}$ (with associated variance $\boldsymbol{\sigma_M^2}$) has been carried out, which informs us on the charge position. Using Bayes theorem, we then write

$$P(\boldsymbol{r}|\boldsymbol{M})P(\boldsymbol{M}) = P(\boldsymbol{r})P(\boldsymbol{M}|\boldsymbol{r}). \qquad (6)$$

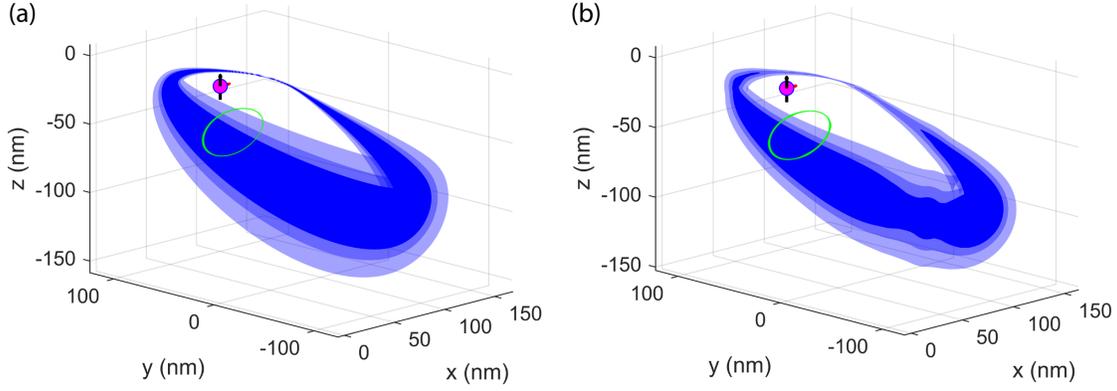

**Supplementary Figure 6:** Probability distribution for NV$_G$ and NV$_D$ (green and blue loops, respectively) as determined from using NV$_F$ as the probe (magenta circle) using a Bayesian approach; the blue arrow indicates the direction of the crystal field at NV$_F$. (b) Same as in (a) but using Gaussian error propagation; the result closely reproduces that in (a) but demands a shorter computing time.

Assuming no prior on the charge position, $P(\boldsymbol{r})$ is a constant thus allowing us to cast $P(\boldsymbol{r}|\boldsymbol{M})$ in the form

$$P(\boldsymbol{r}|\boldsymbol{M}) = \frac{P(\boldsymbol{M}|\boldsymbol{r})}{\iiint P(\boldsymbol{M}|\boldsymbol{r})dr^3}, \qquad (7)$$

where we have used the relation $P(\boldsymbol{M}) = \iiint P(\boldsymbol{M}|\boldsymbol{r}')P(\boldsymbol{r}')dr'^3$ and the integral extends over the diffraction volume. $P(\boldsymbol{M}|\boldsymbol{r})$ can be calculated using the field histogram created by a charge at position $\boldsymbol{r}$ as well as the measurement variance. As an illustration, Supplementary Fig. 6a shows the probability distribution for the positions of NV$_G$ and NV$_D$ as seen by NV$_F$.

The Bayesian approach above is accurate but too resource intensive for cases where many iterations are necessary (the case in Section 7, below). We can gain computational speed via a simpler strategy that relies on summing many Gaussian distributions obtained by the impact of measurement errors at a given point. We can gain computational speed via a simpler strategy that relies on a Gaussian distribution for all errors. Assuming that the presence of a charge $q$ rotates the transverse field by a known angle $\theta_k$, we calculate the probability $P(\boldsymbol{r}|\boldsymbol{M}, \theta_k)$ from the error propagation formula as the 3D Gaussian distribution

$$G_{\theta_k}(\boldsymbol{r}) = \prod_{\nu=x,y,z} \frac{1}{\sigma_\nu \sqrt{2\pi}} \exp\left(-\frac{(r^\nu - R_k^\nu)^2}{2\sigma_\nu^2}\right), \qquad (3)$$

centered at $\boldsymbol{R_k} = \boldsymbol{R}(\theta_k, \mathcal{E}_\parallel - \Upsilon_\parallel, \mathcal{E}_\perp, \Upsilon_\perp)$ with variances defined as

$$\begin{pmatrix} \sigma_x^2 \\ \sigma_y^2 \\ \sigma_z^2 \end{pmatrix} = \left(\frac{\partial \boldsymbol{R}}{\partial \Delta \mathcal{E}_\parallel}\right)^2 \sigma_{\Delta \mathcal{E}_\parallel}^2 + \left(\frac{\partial \boldsymbol{R}}{\partial \mathcal{E}_\perp}\right)^2 \sigma_{\mathcal{E}_\perp}^2 + \left(\frac{\partial \boldsymbol{R}}{\partial \Upsilon_\perp}\right)^2 \sigma_{\Upsilon_\perp}^2. \qquad (4)$$

We then approximate the probability $P(\boldsymbol{r}|\boldsymbol{M})$ in the general case where $\theta$ is not known, by performing a sum of the Gaussian distributions for a series of $\theta$s, i.e.,

$$P(\boldsymbol{r}|\boldsymbol{M}) = \frac{1}{Z}\sum_k z_k\, P(\boldsymbol{r}|\boldsymbol{M}, \theta_k), \qquad (5)$$

where we introduced the renormalization factor $Z = \iiint \sum_k P(\boldsymbol{r}|\boldsymbol{M}, \theta_k)dr^3$ and $z_k = \sigma_{k,x}\sigma_{k,y}\sigma_{k,z}(2\pi)^{\frac{3}{2}}$. The distribution of $\theta_k$s we use is chosen such that adjacent Gaussian probabilities overlap to the same extent by applying the condition $G_{\theta_k}(\boldsymbol{r} = \boldsymbol{R_{k-1}}) \cong 0.8$. Together with the weight $z_k$, the latter ensures that the probability distribution $P(R(\theta, \boldsymbol{M})|\boldsymbol{M})$ is equal for any $\theta$.

Supplementary Fig. 6b shows again the probability distributions for NV$_G$ and NV$_D$ when probed by NV$_F$ as derived from Gaussian error propagation. We find the result closely reproduces that obtained via the Bayesian approach while demanding only a fraction of the computational time. We therefore follow one route or the other depending on the complexity of the problem at hand.

## 7. Drift of the background bias field

The crystal strain and electric field applied to the NV have similar effects[7], here we refer to their added contribution generically as the δ-field. For a repetitive set of experiments, we distinguish the average bias field an NV sees from temporal fluctuations from one spectrum to the next. Consistent with previous observations over longer distances[18,47], we find that NVs tens of nm from each other experience a local bias field different from one another, even for defects along the same crystalline axis. Interestingly, the average bias field of NVs is not constant and can slowly drift over timescale of several hours to days, as shown in Supplementary Fig. 7 for the NV cluster in Fig. 3 of the main text. Alternative mechanisms could be responsible for these drifts. As observed in Monge et al.[18] and consistent with works on bound membranes[48], the diamond attachment method impacts the overall strain in the crystal, implying that minute, slow changes can create drifts. Changes could also be caused by a progressive rearrangement of a metastable electric environment (in the bulk, or on the surface). Indeed, the 100 MHz shift observed here could be induced by a change of one elementary charge at a 150-nm distance, 40 charges at 1 μm, or 1000 charges at 5 μm; such rearrangements have been observed[49]. All field histograms in the manuscript were obtained from spectral series showing no visible drifts. For Fig. 4 of the main text, two datasets obtained months apart and featuring spectral shift differing by hundreds of MHz were used separately to map out all three NVs (and the adjacent dark traps). Importantly, we found the resulting NV localization to be robust from one dataset to another provided we account for a rotation of the transverse field. While these drifts complicate the analysis of field histograms, one could likely apply a correction, or exploit the changing orientation of the transverse field to enhance sensitivity. The same multi-sensor spatial analysis can also be extended to the slower drift to determine their origin.

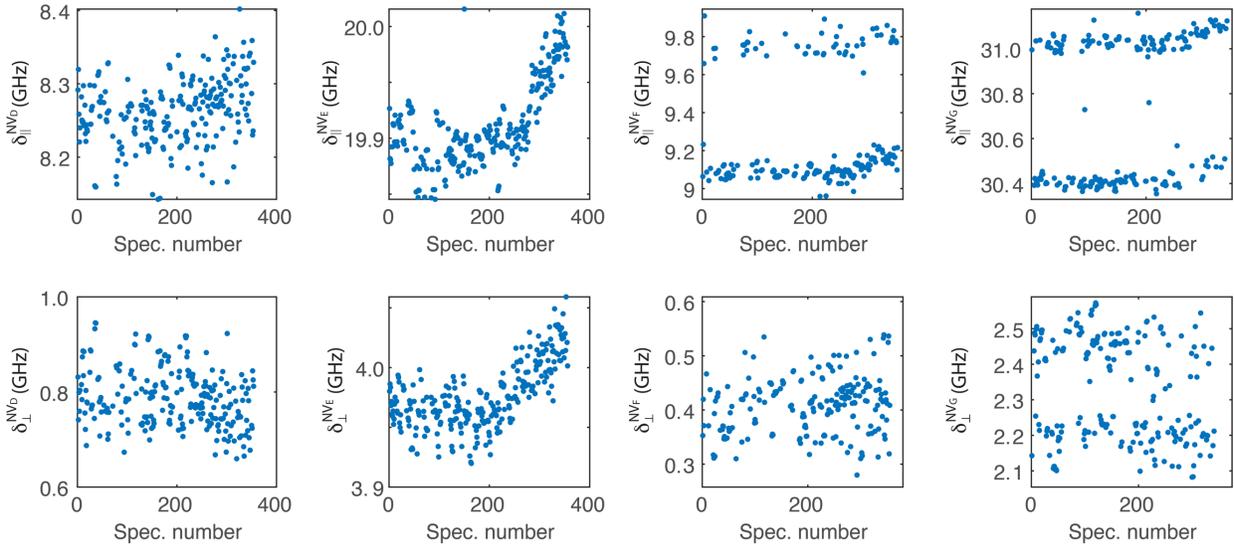

**Supplementary Figure 7:** Longitudinal and transverse components of the electric fields for the NVs in the cluster of Fig. 3 in the main text during a span of 10 days. A 100 MHz shift of the fields is visible for most NVs. Each spectrum takes 25 min, the average interval between spectra is 40 mins.

## 8. Time-correlation of the electric field

At low enough excitation power, we can perform several repump-free PLE sweeps in a row without green re-initialization of the charge environment. This opens the prospect of studying the evolution of the electric environment over time. Here, we recorded a series of 270 spectra under 3 nW of red excitation with one repump every three spectra. Note that under these parameters, the probabilistic ionization of the NVs we read out remains important. This could be overcome by using a combination of higher NA, lower power, lower detector dark count and higher acquisition time. For each three-PLE-sweep set, we track the electric field shift between consecutive spectra as well as the evolution of the charge state of all NVs. Fig. 4b of the main text shows the field shift histograms we measure. We observe this time-evolution of the electric environment under resonant excitation is slow such that the field shift histograms are dominated by NV ionization events. By measuring the field shifts instead of the absolute value of the electric field, we negate background contributions that do not vary, which improves the minimum shift we can measure and increases the sensitivity and range of our sensor.

## 9. Localization using a tri-partite interaction

In general, one can use the multipartite interaction in a cluster of more than two NVs to infer their relative positions and local bias fields. In the present case, we developed an algorithm adapted to 3 NVs, two of whom have the same crystalline axis ($NV_F$ and $NV_G$ in the cluster of Figs. 3 and 4 of the main text). As a brief overview, we first assume $NV_G$ is at the origin, and set its crystalline (bias $\delta$-field) axis along $z$ ($x$). We then proceed to calculate the probability $P(r)$ for the relative positions of $NV_D$ and $NV_F$, respectively $\mathbf{r}_D$ and $\mathbf{r}_F$, as well as the relative angle $\phi_F$ between the transverse bias fields of $NV_F$ and $NV_G$ (i.e., the angle between the average of $\boldsymbol{\delta}_\perp^{NV_F}$ and $\boldsymbol{\delta}_\perp^{NV_G}$). Using the shortened notation $r = (\mathbf{r}_D, \mathbf{r}_F, \phi_F)$, we subsequently assess the probability $P(r)$ by sequentially combining measurements obtained from the field histograms of $NV_F$ and $NV_G$. Throughout our analysis, we use each successive measurement to eliminate unlikely values of $r$ and accelerate the search. The resulting probability density is then projected on the $\mathbf{r}_D$ and $\mathbf{r}_F$ subspaces, yielding probability densities for those two NVs and an average value for $\phi_F = 80 \pm 20°$.

In the following, we refer to the NVs generating and sensing the electric field as the "source" and "probe" NVs, respectively. The steps below describe the search algorithm in further depth.

1 – For all combinations of source and probe NVs, respectively $NV_\alpha$ and $NV_\beta$, $\alpha, \beta = D, E, F, G$, we calculate the probability distribution for $NV_\alpha$ relative to $NV_\beta$ as the sum of Gaussian distributions $\sum_k G_k^{\alpha,\beta}$, where $G_k^{\alpha,\beta}$ is centered at $\mathbf{R}_k^{\alpha,\beta} = \mathbf{R}\left(\theta_k^{\alpha,\beta}, \Delta\varepsilon_\parallel^{\alpha,\beta}, \varepsilon_\perp^{\alpha,\beta}, Y_\perp^{\alpha,\beta}\right)$ with standard deviation $\sigma_k^{\alpha,\beta}$, and the sum extends over a discrete set of $k$ positions along the solution loop in 3D space. We also calculate the electric field $\boldsymbol{\mathcal{E}}_c\left(\alpha, \beta, \theta_k^{\alpha,\beta}\right)$ that $NV_\alpha$ at position $\mathbf{R}_k^{\alpha,\beta}$ would exert on $NV_\beta$.

2 – For each probe $NV_\beta$, and for every pair of source NVs, $NV_\alpha$ and $NV_\gamma$, $\alpha, \beta, \gamma = D, E, F, G$, we iterate through all Gaussian distributions of $G_k^{\alpha,\beta}$ and $G_{k'}^{\gamma,\beta}$ describing the positions of $NV_\alpha$ and $NV_\gamma$ to calculate the transverse electric fields created by only one NV, $\boldsymbol{\mathcal{E}}_{c,\perp}\left(\alpha, \beta, \theta_k^{\alpha,\beta}\right)$ and $\boldsymbol{\mathcal{E}}_{c,\perp}\left(\gamma, \beta, \theta_{k'}^{\gamma,\beta}\right)$, and by the sum of both, $\boldsymbol{\mathcal{E}}_{\Sigma,\perp}$. Since the transverse field is a vector, we have:

$$\left|\boldsymbol{\mathcal{E}}_{\Sigma,\perp}\left(\alpha, \beta, \gamma, \theta_k^{\alpha,\beta}, \theta_{k'}^{\gamma,\beta}\right)\right| = \left|\mathbf{Y}_\perp + \boldsymbol{\mathcal{E}}_{c,\perp}\left(\alpha, \beta, \theta_k^{\alpha,\beta}\right) + \boldsymbol{\mathcal{E}}_{c,\perp}\left(\gamma, \beta, \theta_{k'}^{\gamma,\beta}\right)\right|. \quad (8)$$

We compare the calculated values for $\boldsymbol{\mathcal{E}}_{\Sigma,\perp}$ with that obtained experimentally from the field histogram $\boldsymbol{\mathcal{E}}_{\Sigma,\perp}^{\exp}$, recording the mismatch normalized by the experimental standard deviation, namely, $Z_{\Sigma,\perp}\left(\alpha, \beta, \gamma, \theta_k^{\alpha,\beta}, \theta_{k'}^{\gamma,\beta}\right) = |\boldsymbol{\mathcal{E}}_{\Sigma,\perp}\left(\alpha, \beta, \gamma, \theta_k^{\alpha,\beta}, \theta_{k'}^{\gamma,\beta}\right) - \boldsymbol{\mathcal{E}}_{\Sigma,\perp}^{\exp}|/\sigma_\Sigma$. We then use $Z_{\Sigma,\perp}$ to discard the unlikely $\theta_k^{\alpha,\beta}, \theta_{k'}^{\gamma,\beta}$ pairs (typically $Z_{\Sigma,\perp} > 2$), and subsequently weigh the probability densities of the remaining ones.

3 – We test the most likely relative position between two NVs of the same orientation, $NV_\alpha$ and $NV_\beta$. To do so, we iterate through all Gaussian distributions $G_k^{\alpha,\beta}$ and $G_{k'}^{\beta,\alpha}$. Note that crystallographically equivalent NVs do not necessarily experience local (transverse or longitudinal) δ-fields of the same orientation. We therefore also test the two possible longitudinal strain directions and sweep the angle $\phi_m^{\alpha,\beta}$ between the transverse strain of the two NVs. For every degeneracy of parameters $\mathbf{P} = \theta_k^{\alpha,\beta}, \theta_{k'}^{\gamma,\beta}, \phi_m^{\alpha,\beta}$ we calculate $\mathbf{Z}_{k,k',m}^{\alpha,\beta} = |\mathbf{R}_k^{\alpha,\beta} - \mathbf{O}_{\phi_m^{\alpha,\beta}}(\mathbf{R}_{k'}^{\beta,\alpha})| \cdot \left(\sigma_k^{\alpha,\beta^2} + \mathbf{O}_{\phi_m^{\alpha,\beta}}(\sigma_{k'}^{\beta,\alpha^2})\right)^{-1/2}$ and discard unlikely degeneracies when $\min(\mathbf{Z}_{k,k',m}^{\alpha,\beta}) > 2$. Here, $\mathbf{O}_{\phi_m^{\alpha,\beta}}$ describes the rotation due to the change in transverse strain axis between the two NVs. All remaining pairs of Gaussian distributions are multiplied, i.e., $\mathbf{G}'_{k,k',\phi}^{\alpha,\beta} = \frac{1}{F_{k,k',\phi}} \cdot \mathbf{G}_k^{\alpha,\beta}(\mathbf{r}) \cdot \mathbf{G}_{k'}^{\beta,\alpha}(-\mathbf{r})$, which results in a new normal Gaussian distribution of lower amplitude (with $F_{k,k',\phi}$ the renormalization factor); note this distribution has, in general, a different center and standard deviation, which we denote $\mathbf{R}_{k,k',\phi}^{\alpha,\beta}$ and $\sigma_{k,k',\phi}^{\alpha,\beta}$.

4 – For every parameter set $\theta_k, \theta_{k'}, \phi_m$, we then calculate the position distribution of a third NV in two independent ways: directly relative to $NV_\alpha$, and using the relative positions of $NV_\alpha$ and $NV_\beta$, $\mathbf{R}_{k,k',\phi}^{\alpha,\beta}$, as well as a new distribution of position for $NV_\gamma$ relative to $NV_\beta$. We proceed similarly to step 3, i.e., we discard cases where the mismatch of the two calculated positions is large compared to the combined variances, and multiply the remaining Gaussian distributions to find a new one for the position of $NV_\gamma$ relative to $NV_\beta$. We determine a new renormalization factor for each distribution.

5 – We sum all Gaussian distributions for the positions of $NV_\alpha$ and $NV_\gamma$ relative to $NV_\beta$, weighing each with their two renormalization factors. We finally renormalize each of the two distributions for $NV_\alpha$ and $NV_\gamma$.

## 10. Co-localization of dark charge traps

As seen in Fig. 2, the NV cluster in Fig. 3 of the main text also contains non-fluorescent (i.e., "dark") charge traps. In an extended dataset partially shown in Supplementary Fig. 8, we observe discrete shifts of the resonances of both $NV_F$ and $NV_G$, which we attribute to carrier capture by a dark trap $DT_A$. These shifts occur 5 times in a series of 225 spectra (2.2±1%). Note that we observed a distinct, stronger shift with similar occurrences that could not be analyzed due to the limited range of our laser sweeps. In addition, the field histogram of $NV_G$ displays discrete jumps that do not correlate with the charge state of any other NV and that we attribute to carrier capture by a second dark trap, $DT_B$. On average, these shifts occur in 19% of all spectra (13% and 25% when $NV_F$ is in its negative and neutral state, respectively). Once a trap is identified, we calculate the probability density for its location relative to $NV_F$ and $NV_G$ assuming a given charge (one electron or one hole). We then use our knowledge of the position and strain of the two NVs to combine that information by simply multiplying and renormalizing the two probability densities (Figs. 4d and 4e in the main text). The relative quality of the intersection of the two loops informs us on the charge sign: By comparing the renormalization factor for a negative and positive charge trap, we find $DT_A$ ($DT_B$) is $10^{300}$ ($10^3$) more likely to have one extra electron 2.2% (19%) of the time compared to one hole.

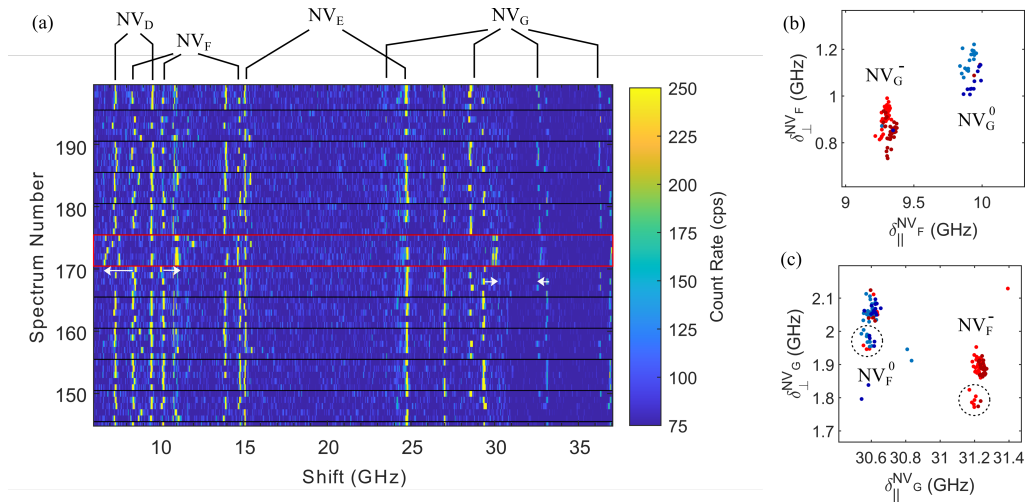

**Supplementary Figure 8:** (a) Recursive optical spectroscopy of the NV cluster in Fig. 3 of the main text from an extended dataset with a modified strain. We identify discrete shifts of $NV_F$ and $NV_G$ resonances for the five spectra in the red square, as highlighted by the white arrows. (b) Field histograms for $NV_F$. Red and blue dots indicate the charge state of $NV_G$, whereas light and dark shades label the charge state of $NV_D$. (c) Same as in (b) but after exchanging the roles of $NV_G$ and $NV_F$. While the shift due to the $DT_A$ is out of range, we find a discrete shift of the electric at $NV_G$, which we attribute to a second dark trap, $DT_B$ (dashed circles). The green (red) laser power during charge initialization (frequency sweep) is 1 μW (25 nW); a 10-s, 200-μW green pulse is used for alignment every 5 spectra (black lines in (a)) and the reference frequency is 470.470 GHz.